\documentclass{elsart}

\usepackage{indentfirst}
\usepackage{graphicx}
\usepackage{psfrag}

\usepackage{subfigure}

\usepackage[english]{babel}
\usepackage[latin1]{inputenc}

\usepackage[sort,sectionbib]{natbib}

\usepackage{xspace} 
\usepackage{amsmath,amssymb}

\usepackage{color}
\definecolor{oneblue}{rgb}{0,0.0,0.75}
\usepackage[colorlinks,
            urlcolor=oneblue,
            linkcolor=oneblue,
            citecolor=oneblue,
            bookmarksopen=false,
            pagebackref]{hyperref}

\newtheorem{definition}{Definition}

\newcommand{\pd}[2]{\frac{\partial#1}{\partial#2}}

\def\u{\vec{u}}
\def\x{\vec{x}}
\def\b{\vec{b}}

\def\n{\vec{n}}
\def\f{\vec{f}}
\def\v{\vec{v}}
\def\0{\vec{0}}
\def\S{\mathcal{S}}
\def\div{\nabla\cdot}
\def\I{\underline{\underline{I}}}
\def\eps{\underline{\underline{\varepsilon}}}
\def\ssigma{\underline{\underline{\sigma}}}

\journal{Computer Methods in Applied Mechanics and Engineering}

\begin{document}

\begin{frontmatter}

\title{Influence of sedimentary layering on tsunami generation}
\author[Dutykh]{Denys Dutykh}
\address[Dutykh]{CMLA, ENS Cachan, CNRS, PRES UniverSud, 61 Av. President Wilson, F-94230 Cachan, FRANCE -
Now at Université de Savoie, Laboratoire de Mathématiques LAMA - UMR 5127,
Campus Scientifique, 73376 Le Bourget-du-Lac, FRANCE}
\ead{Denys.Dutykh@univ-savoie.fr}

\author[Dias]{Fr\'{e}d\'{e}ric Dias\corauthref{cor}}
\address[Dias]{CMLA, ENS Cachan, CNRS, PRES UniverSud, 61 Av. President Wilson, F-94230 Cachan, FRANCE}
\corauth[cor]{Corresponding author.}
\ead{Frederic.Dias@cmla.ens-cachan.fr}

\begin{abstract}
The present article is devoted to the influence of sediment layers on the process of tsunami generation. The main scope here is to demonstrate and especially quantify the effect of sedimentation on vertical displacements of the seabed due to an underwater earthquake. The fault is modelled as a Volterra-type dislocation in an elastic half-space. The elastodynamics equations are integrated with a finite element method. A comparison between two cases is performed. The first one corresponds to the classical situation of an elastic homogeneous and isotropic half-space, which is traditionally used for the generation of tsunamis. The second test case takes into account the presence of a sediment layer separating the oceanic column from the hard rock. Some important differences are revealed. We conjecture that deformations in the generation region may be amplified by sedimentary deposits, at least for some parameter values.
The mechanism of amplification is studied through careful numerical simulations.
\end{abstract}


\end{frontmatter}


\section{Introduction}

The primary application of this study is that of tsunami generation by the deformation of the sea bottom following an underwater earthquake. We do not explicitly compute the tsunami waves induced above the generation region. The coupling between solid and water motions was already performed in our previous work \citep{DD09} and can be done again if necessary. Here we are mainly interested in the extreme amplitudes of the seabed displacements during the first minutes of a tsunamigenic earthquake. Recall that the free surface motion roughly follows these displacements. There are two fundamental reasons for this. The first one is that the rupture velocity of the seismic source, $V$, is much larger than the phase
velocity of the tsunami, $c$. In practice, for seismic sources, $V$ is of the order of 3 km/s, whereas $c$ is typically less than 250 m/s, even for the deepest ocean basins \citep{OkalSyno2003}. It means that the gravitational forces do not have enough time to change the shape of the free surface during the characteristic time of the seabed motion \citep{Ben-M}. The second reason is that water is assumed to be incompressible and shallow. Altogether it means that for our purpose we can restrict our attention to the motion of the ocean bottom. Profiles of the ocean free surface are not computed in this paper.

The two fundamental reasons mentioned above are often used to justify the passive approach for tsunami generation where the static sea-bed displacement is simply translated to the free surface to generate the initial condition. Our previous investigations \citep{ddk, Dutykh2006, DD09} showed important differences between passive and active generations when the resulting wave is generated by a moving bottom.

We would like to underline that the present study is a theoretical one at this stage and must be followed by a thorough study of the influence of the various parameters such as rupture velocity, focal depth, dip angle, slip angle, fault size (only one-dimensional -- 1D -- faults are considered in this paper), sedimentary structure (the sedimentary layer is taken as a Poisson solid in this paper). Moreover, the far-field effect might be different from the near-field effect. We simply explore some aspects of the influence of the sedimentary layering on tsunami generation. We do not consider historical examples, even though it should be done in the future. Our goal is to present a framework for studying the process of tsunami generation. Recall that
ten years ago, Synolakis et al. \citep{SLCY1997} were writing: ``There is a lack of quantitative information on sediment
layers overlying tsunamigenic faults and about how these layers affect directly the generation of tsunamis.'' Our study
is a small step toward a better understanding of the r\^ole of sediments. 

The influence of sedimentary layering was already mentioned in some studies \citep{Fukao1979, Okal1988, Fuller2006}. Let us comment on the various results obtained so far. Both studies \citep{Fukao1979} and \citep{Okal1988} point out that fracturing through thick sediments produces large displacements in the source region but relatively small displacements in the far field. In the work by Okal \citep{Okal1988}, the influence of the sediment layer was studied in the framework of normal modes and interesting results were obtained for sources inside as well as outside the layer of sediments. In the present study we perform direct numerical simulations by solving the elastodynamics equations with a finite element method (FEM). 

To our knowledge, the most recent numerical study concerning the r\^ole of sedimentation in subduction-zone thrust faults is \citep{Fuller2006}. The scope of that paper was the long-term evolution of a typical subduction wedge. A quite sophisticated thermo-mechanical modelling of the plate movement with realistic rheology was used. As in our study, the governing equations were solved with a two-dimensional FEM. The authors came to some important conclusions  (see also \citep{Syno2006} for interesting remarks on sediment layers). We would like to quote some of them since there is a connection with our results:
\begin{quote}
``Our numerical simulations demonstrate that sedimentation stabilizes the underlying wedge, preventing internal deformation beneath the basin. Maximum slip during great-thrust earthquakes tends to occur where sedimentary basins stabilize the overlaying wedge. The lack of deformation in these stable regions increases the likelihood of thermal pressurization of the subduction thrust, allows the fault to load faster, and allows greater healing of the fault between rupture events.''
\end{quote}

\begin{figure}
	\centering
		\includegraphics[width=0.99\textwidth]{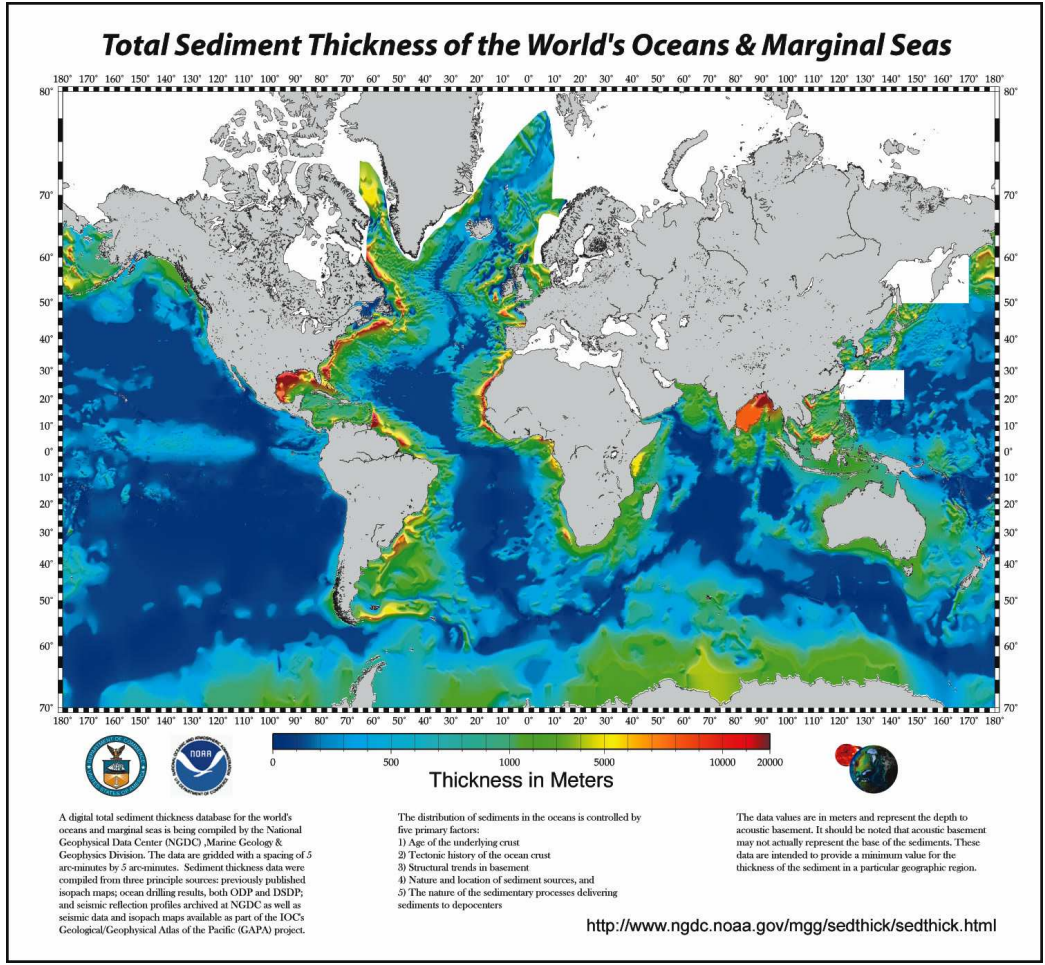}
	\caption{Total sediment thickness of the world's ocean and marginal seas (source: NOAA).}
	\label{fig:sedthick9}
\end{figure}

\begin{figure}
	\centering
		\includegraphics[width=0.99\textwidth]{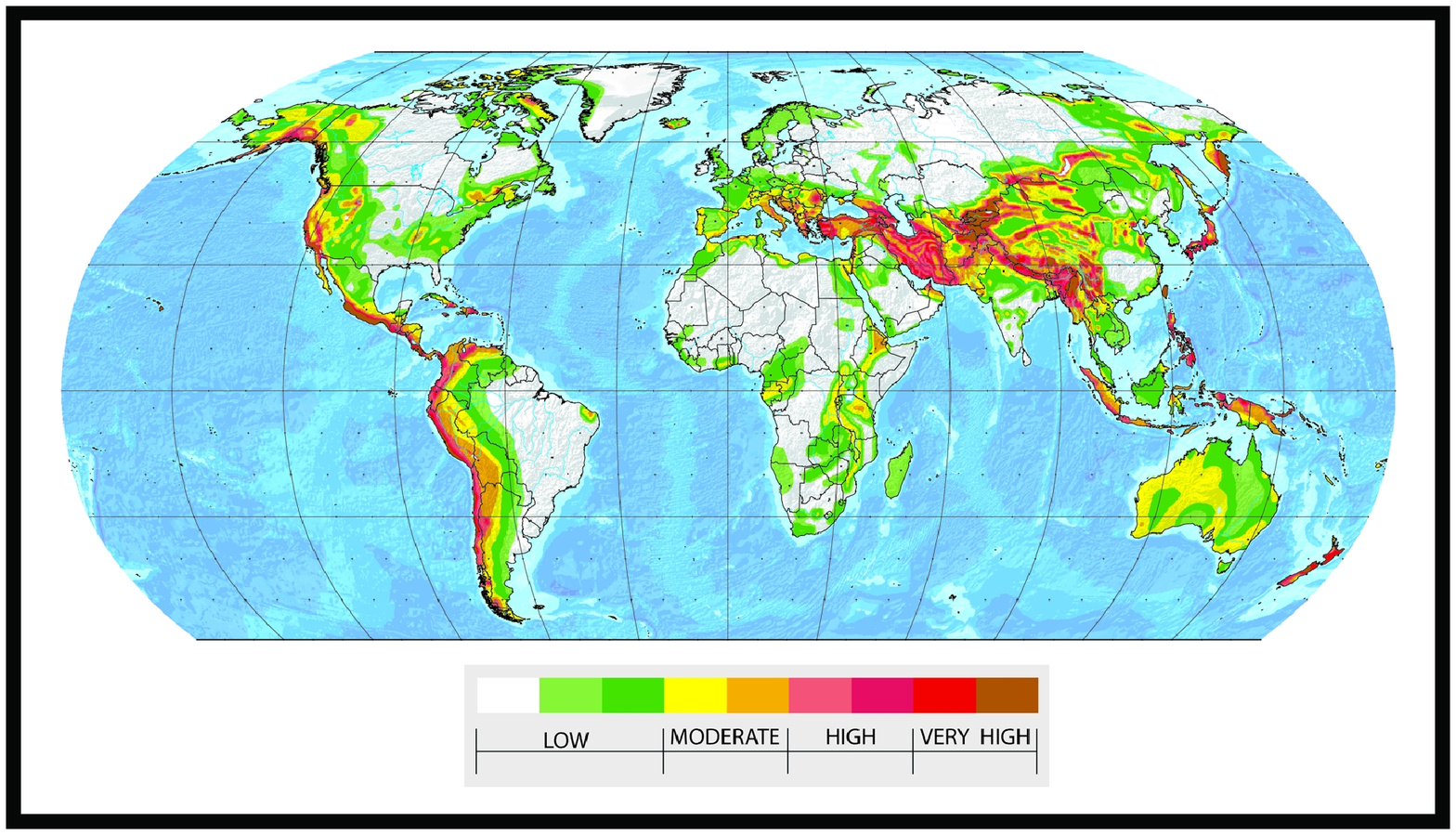}
	\caption{Global seismic hazard map (source: Swiss Seismological Service).}
	\label{fig:global_seismic_hazard}
\end{figure}

In view of the above results, it is interesting to compare the distribution of sediment thickness in the world oceans (see \figurename~\ref{fig:sedthick9}) with the seismic hazard map (see \figurename~\ref{fig:global_seismic_hazard}). As one can see on \figurename~\ref{fig:sedthick9}, the sediment thickness varies from 0 to 20 km. Important deposits can be found along the Eastern coasts of America and the Western coasts of Africa. Fortunately there is no substantial seismic activity in these regions. But in the Bay of Bengal, and in particular in the Andaman sea, the situation is different. In this part of the world two factors are present simultaneously: an important seismic hazard and thick sediment layers. Unfortunately, we do not have reliable information on sediment thickness in the Mediterranean region. One can find different information in the literature. The estimates go from 25 m \citep{Hoogakker2004} to 1500 m or even more \citep{Ergun2005}. So, it is difficult to draw any conclusions. In this study we try to understand what kind of implications this may have for tsunami generation processes.

The present paper is organised as follows. In Section \ref{sec:model} we briefly describe the simplified mathematical model which represents the Earth crust. In the same section we also give some ideas about the discretisation procedure of the governing equations. Section \ref{sec:tests} contains the description of two idealised test cases. Then, some results on the profile
of the seafloor are presented. Finally, important conclusions and practical recommendations to tsunami wave modelers are given in Section \ref{sec:conclusions}. Some directions for future research are also outlined.


\section{Mathematical model and numerical method}\label{sec:model}

In this paper we use the same mathematical model as in our previous study \citep{DD09}. Nevertheless, we give here a brief description of the model and refer to \citep{Dutykh2007a, DD09, KDID} for more details. We emphasize again that we consider a 1D fault. In other words the fault has only one dimension, its length $L$.

The fault is assumed to lie inside a linear elastic isotropic material. In the next section both homogeneous and inhomogeneous distributions of the Earth crust properties will be considered. 

We provide the 3D version of the governing equations. Let $\ssigma$ represent the stress tensor. The displacement field 
$\u = (u,v,w)(x,y,z,t)$ satisfies the classical elastodynamic equations issued from continuum mechanics \citep{Aki2002}:
\begin{equation}
  \div\ssigma = \rho\pd{^2\u}{t^2},
\end{equation}
where $\rho$ the material density. It is common in seismology to assume that the stress tensor is determined by Hooke's law through the strain tensor $\eps = \frac12\left(\nabla\u + \nabla^t\u\right)$. Therefore
\begin{equation}
  \ssigma = \lambda(\div\u)\I + 2\mu\eps,
\end{equation}
where $\lambda$ and $\mu$ are the Lam\'e coefficients. The coefficient $\mu$ is the shear modulus. Thus, we come to the following linear elastodynamic problem:
\begin{equation}\label{eq:dynelast}
  \div\left(\lambda(\div\u)\I + \mu(\nabla\u+\nabla^t\u)\right) = \rho\pd{^2\u}{t^2}.
\end{equation}
As already pointed out, these coefficients can possibly depend on the spatial coordinates $(x,y,z)$ ($x,y$: horizontal, $z$: vertical). The Lam\'e coefficients can be expressed in terms of Poisson's ratio $\nu$ and Young's modulus $E$ as follows:
\begin{equation*}
  \lambda = \frac{2\mu\nu}{1-2\nu} = \frac{E\nu}{(1+\nu)(1-2\nu)}, \quad
  \mu = \frac{E}{2(1+\nu)}.
\end{equation*}
This remark immediately leads to the following property of solutions to the steady version of Equation (\ref{eq:dynelast}): such solutions do not depend on Young's modulus but only on Poisson's ratio.

The fault is modeled as a dislocation inside an elastic material. This type of model is widely used for the interpretation of seismic motion. A dislocation is considered as a surface (in three-dimensional problems) or a line (in two-dimensional problems) in a continuous medium where the displacement field is discontinuous. The displacement vector is increased by the amount of the Burgers vector $\b$ along any contour $C$ enclosing the dislocation surface (or line), i.e.
\begin{equation}
  \oint\limits_{C} d\u = \b.
\end{equation}
We let a dislocation run at speed $V$ along a fault inclined at an angle $\delta$ with respect to the horizontal. The rupture starts at the point $x=0$ and $z=-d$ (it is supposed to be infinitely long in the transverse $y-$direction), propagates at constant rupture speed $V$ for a finite time $L/V$ in the direction $\delta$ and stops at a distance $L$. Let $\zeta$ be a coordinate along the dislocation line. On the fault located in the interval $0 < \zeta < L$, the slip is assumed to be constant. The rise time is assumed to be $0$. We insist here that in real situations, the fault is two-dimensional and the rupture velocity usually refers to the velocity at which the fault ruptures in the transverse $y-$direction (see for example \cite{Okal1988}).

\subsection{Discretization of the elastodynamics equations}

In order to apply the FEM we rewrite the governing equation (\ref{eq:dynelast}) in the domain $\Omega$ in variational form. One has
\begin{equation*}
  \int\limits_{\Omega} \rho\pd{^2\u}{t^2}\cdot\v \;\; d\Omega + 
  \int\limits_{\Omega} \ssigma(\u) :\nabla\v \;\; d\Omega = 
  \int\limits_{\Gamma_N} \f\cdot\v \;\; dS,
  \quad \forall \v \in \mathcal{V},
\end{equation*}
where $\mathcal{V}$ is the linear closed subspace of $\bigl(H^1(\Omega)\bigr)^2$ and $\f$ is the loading applied to the Neumann boundary $\Gamma_N$. This term is equal to zero in our computations, since the seabed is considered to be a free boundary in geophysics.

In order to discretize the time derivative operator we apply a classical second order finite-difference scheme. We underline that the resulting method is fully implicit and has the advantage of being free of any CFL-type condition. In such problems implicit schemes become 
advantageous since the velocity of propagation of seismic waves is of the order of $3 - 4$ km/s. After discretizing in time, one obtains the following variational form:
\begin{multline*}
  \int\limits_{\Omega} 
  \rho\frac{\u^{(n+1)} - 2\u^{(n)} + \u^{(n-1)}}{\Delta t^2}\cdot\v \;\; d\Omega + 
  \int\limits_{\Omega} \ssigma(\u^{(n+1)}) :\nabla\v \;\; d\Omega = 
  \int\limits_{\Gamma_N} \f\cdot\v \;\; dS,
\end{multline*}
where the superscript denotes the time step number, e.g. $\u^{(n)} = \u(\x,t_n)$. Then, we apply the usual $\mathbb{P}2$ finite-element discretization procedure. For the numerical computations, we used the freely available code FreeFem++ \cite{Hecht1998}.

\begin{figure}
	\centering
		\includegraphics[width=0.99\textwidth]{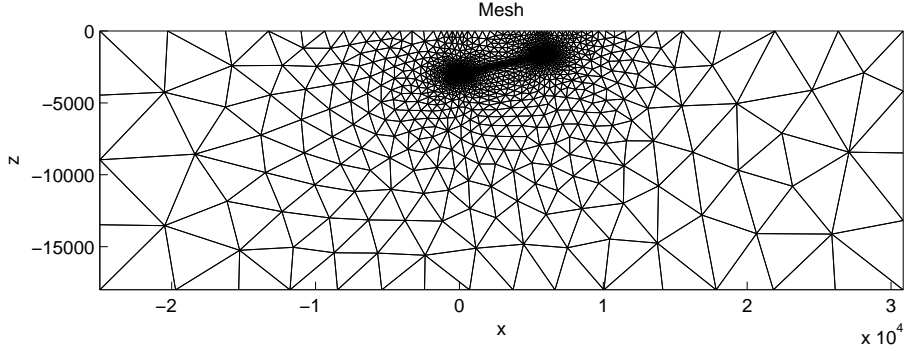}
	\caption{Typical mesh used in the numerical computations. The length scale is in meters.}
	\label{fig:mesh}
\end{figure}

Let us say a few words about the boundary conditions and the treatment of the dislocation in the program. As already stated, the seabed is assumed to be a free surface:
$$
	\ssigma\cdot\n = \f = \vec{0}, \quad z = 0.
$$ 
The other boundaries are assumed to be fixed or in other words, we apply Dirichlet type boundary conditions $\u = \vec{0}$. The authors are aware of the reflective properties of this type of boundary conditions. But we take a computational domain which is sufficiently large, so that the seismic waves do not reach the boundaries during the simulation time. This approach is not computationally expensive since we use adaptive mesh algorithms \cite{Hecht1998} and in the regions far from the fault, element sizes are considerably bigger than in the fault vicinity. A typical mesh used in simulations is plotted on Figure \ref{fig:mesh}.

Now, let us discuss the implementation of the dislocation. Across the fault, the displacement field is discontinuous and satisfies the following relation:
\begin{equation}\label{eq:condburgers}
  \u^+(\x,t) - \u^-(\x,t) = \b(\x,t),
\end{equation}
where the signs $\pm$ denote the upper and lower boundary of the dislocation surface, respectively. In order to satisfy the condition (\ref{eq:condburgers}) we apply the following boundary conditions on the fault surface:
\begin{equation*}
  \left.\u\right|_{\x\in\Gamma^+} = \frac{\b}{2}, \qquad
  \left.\u\right|_{\x\in\Gamma^-} = -\frac{\b}{2}.
\end{equation*}

\textbf{Remark:} Due to the presence of huge hydrostatic pressures in the crust, the two sides of the fault cannot detach physically. In any case this situation does not occur in nature. Mathematically it means that the Burgers vector $\b$ is tangent to the dislocation surface at each point.


\section{Numerical results}\label{sec:tests}

\begin{figure}
			\centering
			  \psfrag{d}{{\tiny $d$}}
			  \psfrag{L}{{\tiny $L$}}
			  \psfrag{O}{{\tiny $O$}}
			  \psfrag{H}{{\tiny $h_s$}}
			  \psfrag{x}{{\tiny $x$}}
			  \psfrag{y}{{\tiny $z$}}
				\includegraphics[width=0.67\textwidth]{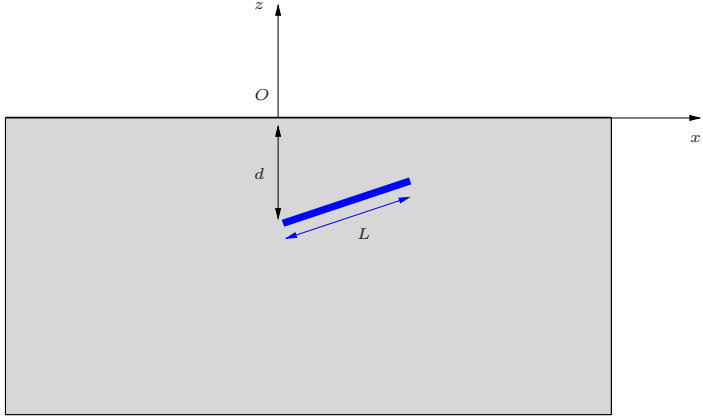}
				\caption{Test case with a homogeneous medium.}
				\label{fig:homogeneous}
\end{figure}

The numerical method described in the previous section was validated in our previous study \citep{DD09} by comparing
numerical results with analytical results.
In the present work we compare vertical displacements at the ocean bottom in two different situations. The first test case corresponds to the traditional modelling procedure where the Earth crust is assumed to be a homogeneous elastic material. It is schematically depicted on \figurename~\ref{fig:homogeneous}. For example, the well known Okada solution\footnote{The original paper by Okada was published in 1985. In the Russian literature this solution was already known in 1978, after the publication of results by Gusiakov \citep{Gusiakov1978}. Some particular cases of Okada solution were known even earlier \citep{Mansinha1971, Freund1976}.} \citep{Okada85}, which is still widely used to construct initial conditions for various tsunami propagation codes, is based on these assumptions. 

\begin{figure}
			\centering
			  \psfrag{d}{{\tiny $d$}}
			  \psfrag{L}{{\tiny $L$}}
			  \psfrag{O}{{\tiny $O$}}
			  \psfrag{H}{{\tiny $h_s$}}
			  \psfrag{x}{{\tiny $x$}}
			  \psfrag{y}{{\tiny $z$}}
				\includegraphics[width=0.67\textwidth]{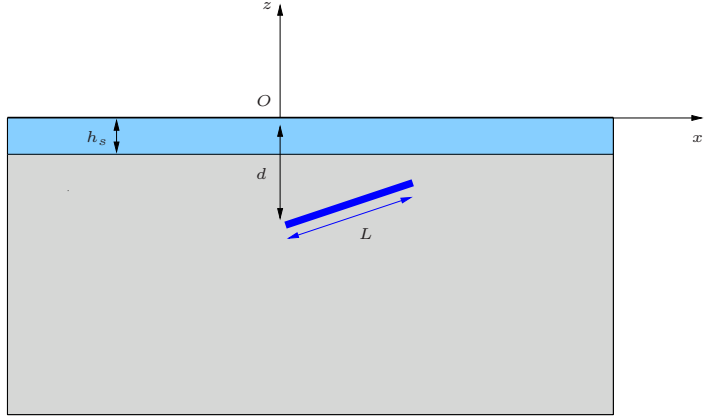}
				\caption{Test case with a sediment layer in between the oceanic column and the hard rock.}
				\label{fig:sand}
\end{figure}

In the second test case, we add a sediment layer of thickness $h_s$ on top of the previous configuration. This situation is depicted on \figurename~\ref{fig:sand}. Let us provide some comments on the model for sediments chosen in this study. In fact, what we call sediments here is an elastic layer which has the mechanical properties of sand according to \citep{Mei1994}. It means that porosity and the two-phase nature of this medium are neglected. These effects should be investigated in the future. As stated by Okal \cite{Okal1988}, detailed oceanic models of sedimentary structure have evidenced strong gradients of seismic velocities. The chosen sedimentary model is not extreme; much looser and weaker structures are encountered in nature.

\begin{table}
     \begin{center}
       \begin{tabular}{c|l}
         \hline\hline
         \textit{\bf parameter} & \textit{\bf value} \\
         \hline
         Fault depth, $d$ & $4 000$ m \\
         \hline
         Dip angle, $\delta$ & $13^\circ$ \\
         \hline
         Fault length, $L$ & $2 000$ m \\
         \hline
         Slip along the fault, $b$ & $10$ m \\
         \hline
         Fault propagation velocity, $V$ & $2 500$ m/s \\
         \hline\hline
       \end{tabular}
       \caption{Values of fault parameters used in this study.}
       \label{tab:constants}
     \end{center}
\end{table}

\begin{table}
     \begin{center}
       \begin{tabular}{c|l|l}
         \hline\hline
         \textit{\bf parameter} & \textit{\bf sand} & \textit{\bf granite} \\
         \hline
         Shear modulus $\mu$, Pa & $2\times10^8$ & $30\times10^9$ \\
         \hline
         Poisson's ratio & $0.3$ & $0.27$ \\
         \hline
         Shear wave velocity $\sqrt{\mu/\rho}$, m/s & $330$ & $3230$ \\
         \hline\hline
       \end{tabular}
       \caption{Values of mechanical parameters for sand and granite.}
       \label{tab:constants2}
     \end{center}
\end{table}

\begin{figure}
		\centering
			\includegraphics[width=0.80\textwidth]{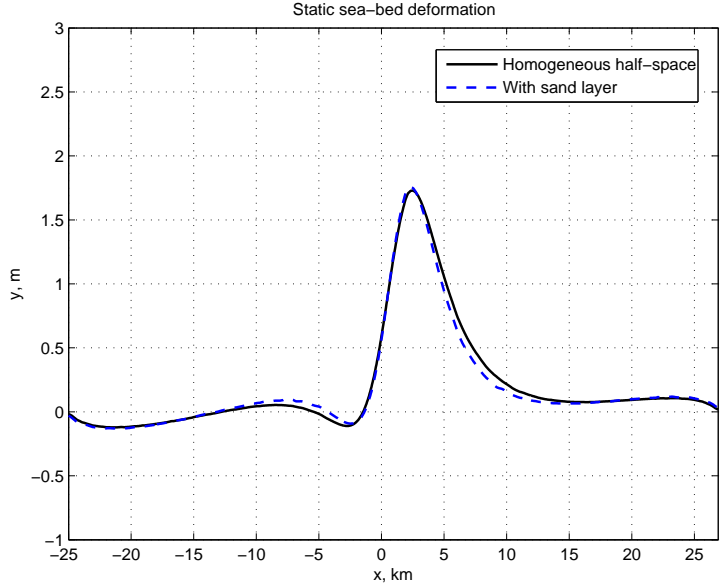}
			\caption{Volterra dislocation source. Static solutions with (dashed line) and without (solid line) sediments. The thickness of the sediment layer is $h_s = 600$ m.}
		  \label{fig:static}
\end{figure}

Let us now discuss the results. First we present static solutions corresponding to the test cases described above. Two solutions are plotted on \figurename~\ref{fig:static}. In this case we take a sediment layer thickness $h_s$ equal to $600$ m. In all figures of this section we plot the vertical displacement at the free surface of the Earth crust (or at the seabed, in other words). The values of the other parameters used in the computations are given in Tables \ref{tab:constants} and \ref{tab:constants2}. As the reader can see, there is no significant difference between the two solutions. In other words, sediments do not influence the static deformations due to a dislocation source. Physically, we can understand this situation, since in the static case the sand layer is just raised up by the deformed granite. Note that the parameters used here are not realistic. Indeed the length of the fault $L$ is quite small and we repeat that the fault is 1D. The slip $b$ is extremely large but since at this stage the problem is linear, the absolute value of the slip is not that important.

There is another rather mathematical explanation. In fact, as mentioned above, the steady Lam\'e equations\footnote{We assume that we neglect volume forces as well. So, the system of governing equations is homogeneous.} do not depend on Young's modulus but only on Poisson's ratio. Not surprisingly the Okada solution has the same property since the analytical expressions contain the Lam\'e coefficients in the combination $\lambda/(\lambda+\mu)$, which depends only on Poisson's ratio $\nu$. One can see in Table \ref{tab:constants2} that Poisson's ratio $\nu$ is almost the same for sand and granite. This is why the sediment layer does not have a strong effect on the steady solution.

\begin{figure}
  \centering
  \subfigure[$t = 1.0 s$]%
  {\includegraphics[width=0.49\textwidth]{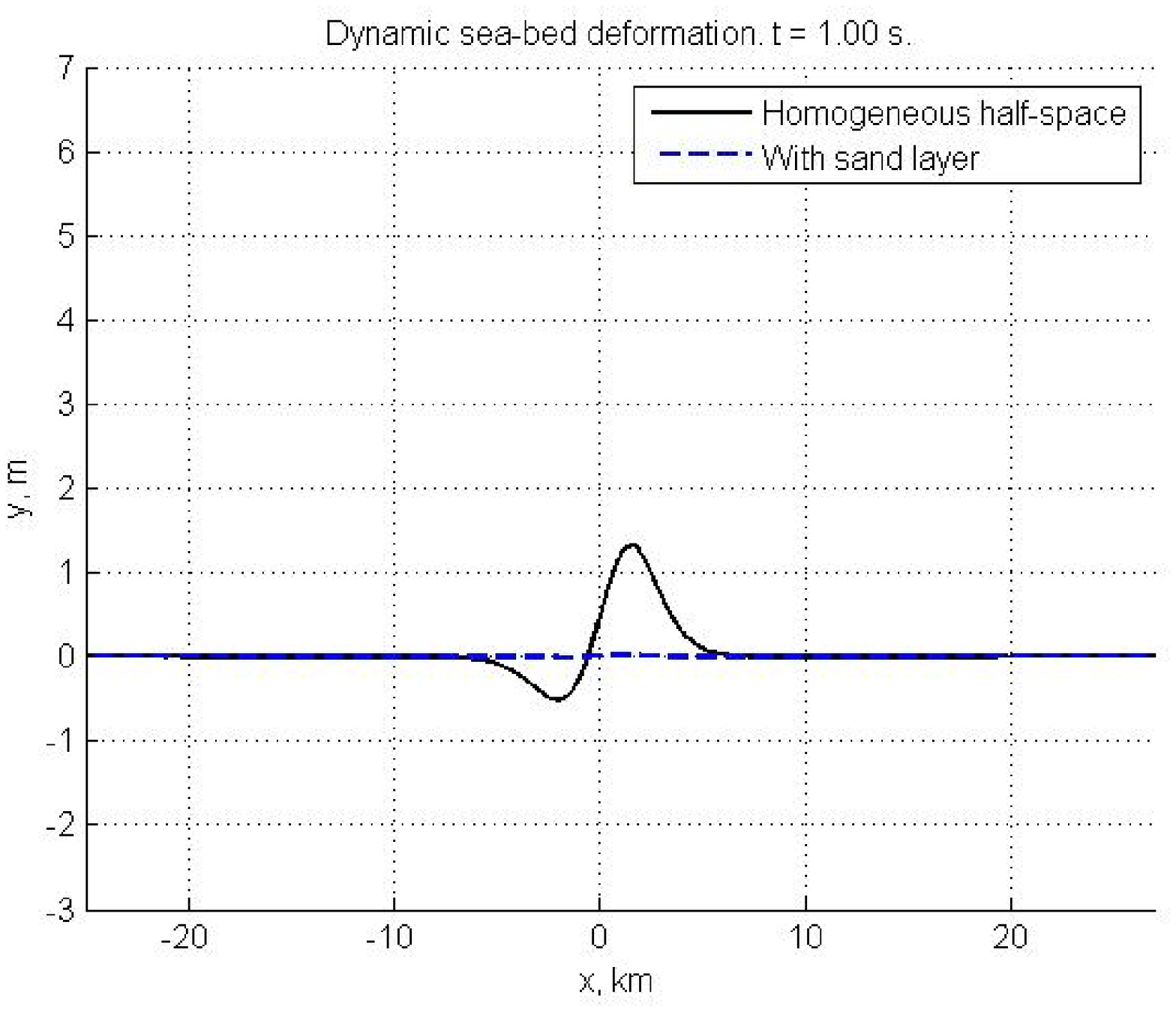}}
  \subfigure[$t = 1.5 s$]%
  {\includegraphics[width=0.49\textwidth]{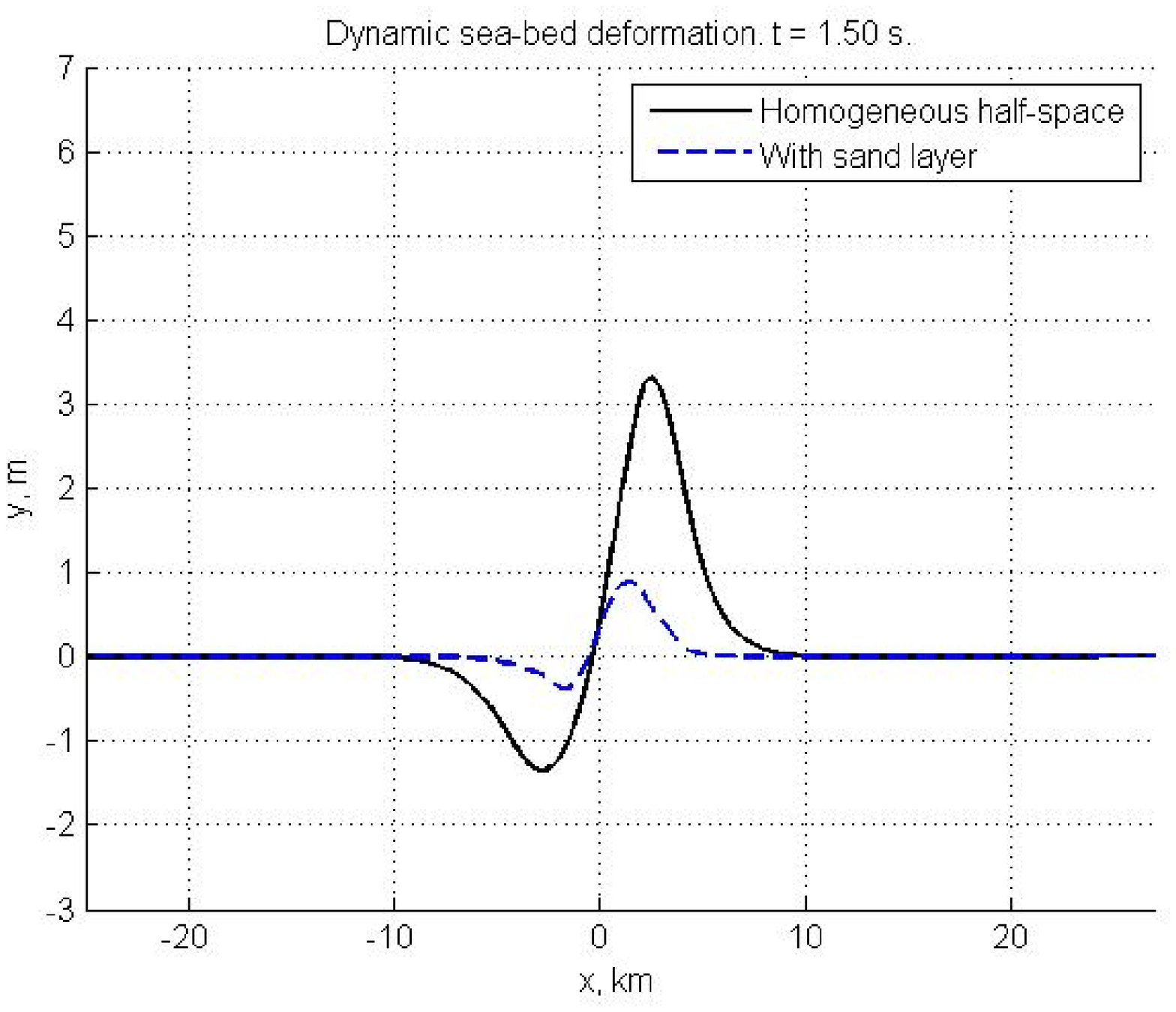}}
  \caption{Dynamic sea-bed displacements at the beginning of the rupture process. The thickness of the sediment layer is $h_s = 600$ m.}
  \label{fig:beginH600}
\end{figure}

\begin{figure}
  \centering
  \subfigure[$t = 2.0 s$]%
  {\includegraphics[width=0.49\textwidth]{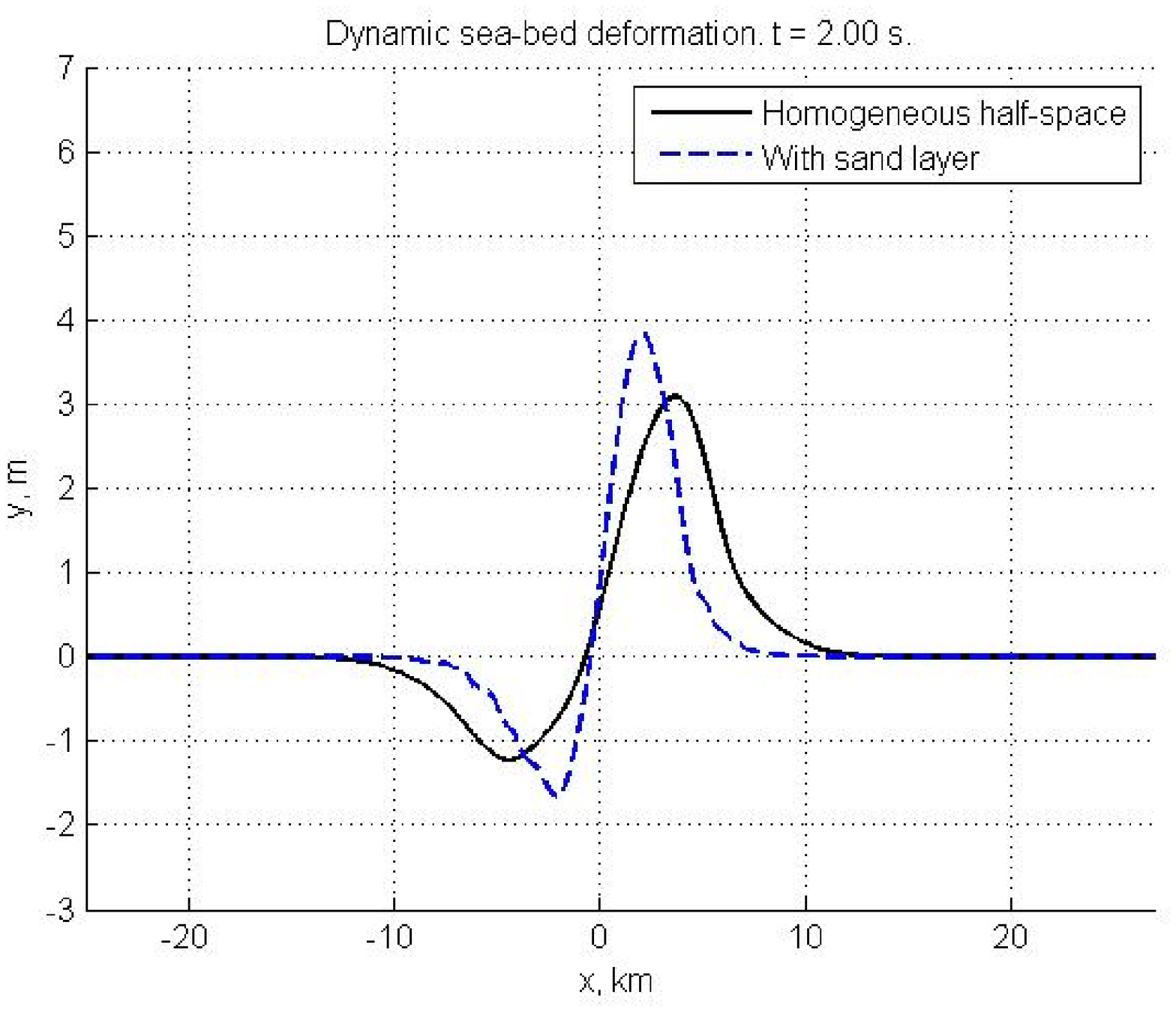}}
  \subfigure[$t = 2.5 s$]%
  {\includegraphics[width=0.49\textwidth]{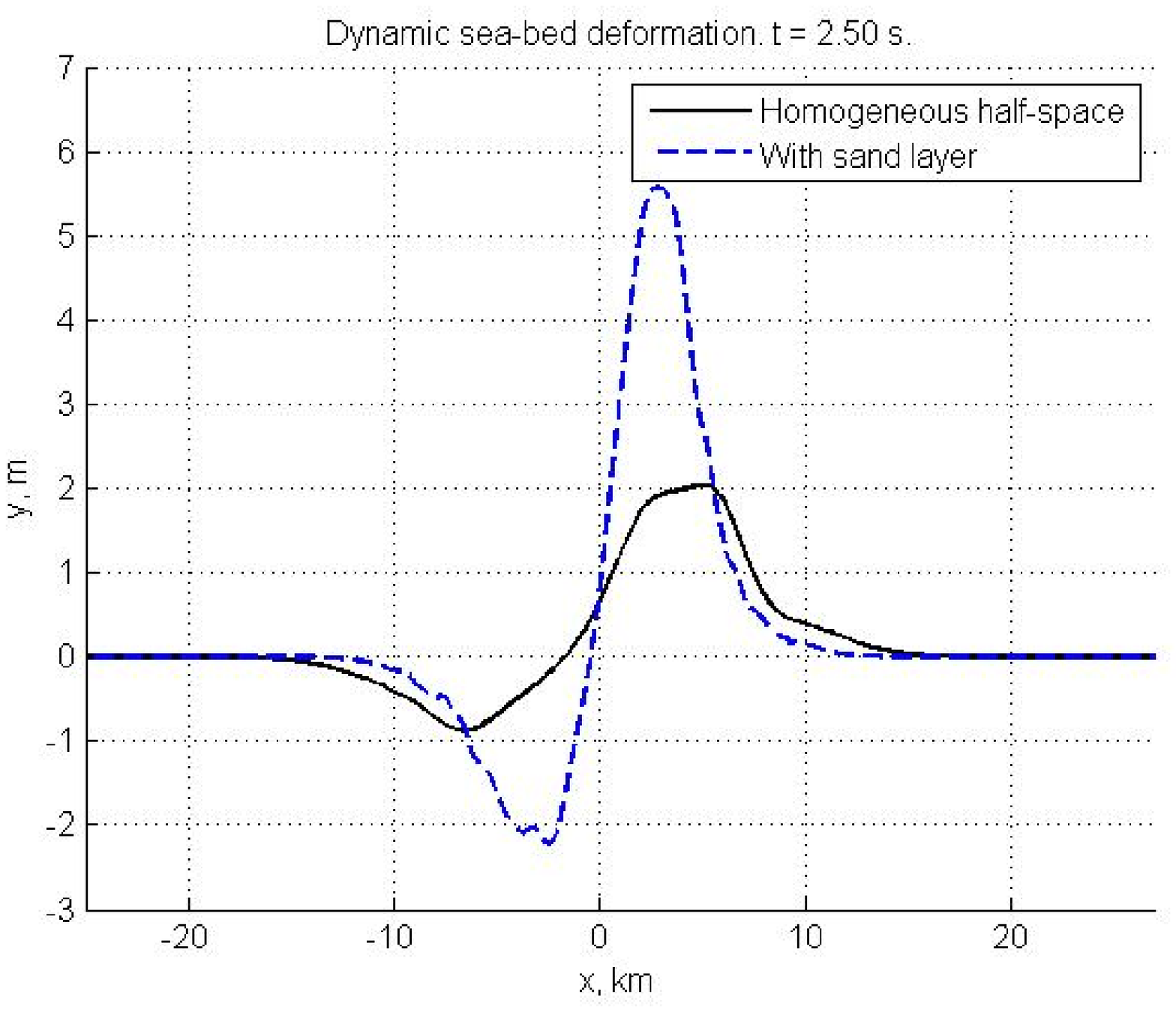}}
  \caption{Dynamic sea-bed displacements. The solution, which takes into account the presence of the sediments, produces much bigger vertical displacements. The thickness of the sediment layer is $h_s = 600$ m.}
  \label{fig:ampl}
\end{figure}

\begin{figure}
  \centering
  \subfigure[$t = 3.0 s$]%
  {\includegraphics[width=0.49\textwidth]{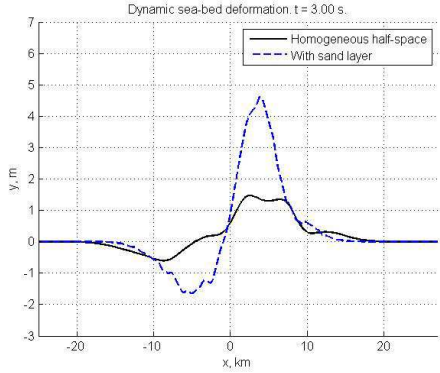}}
  \subfigure[$t = 3.5 s$]%
  {\includegraphics[width=0.49\textwidth]{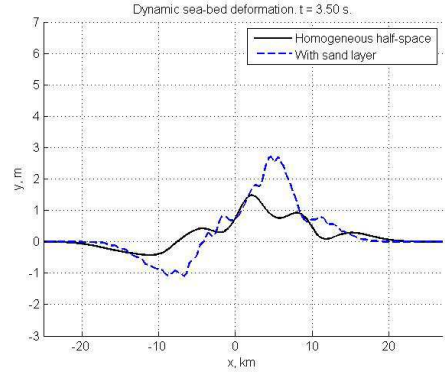}}
  \caption{Dynamic sea-bed displacements at the end of the simulation. The thickness of the sediment layer is $h_s = 600$ m.}
  \label{fig:endH600}
\end{figure}

Now, let us consider the dynamical effects. The results are presented on Figures \ref{fig:beginH600}--\ref{fig:endH600}. In these computations we take the same thickness of the sand layer as in the static case ($h_s = 600$ m). It can be seen on \figurename~\ref{fig:beginH600} that the deformation in the homogeneous case is initiated earlier. This is to be expected since the shear wave velocity in the sand is almost ten times slower than in the granite. Later, the deformation in the inhomogeneous case starts to evolve. It is surprising that it produces much bigger displacements -- see \figurename~\ref{fig:ampl}. In other words, taking into account the sediments increases considerably the seabed deformations. When time evolves, both solutions eventually reach comparable amplitudes (see \figurename~\ref{fig:endH600}).

\begin{figure}
  \centering
  \subfigure[$t = 1.0 s$]%
  {\includegraphics[width=0.49\textwidth]{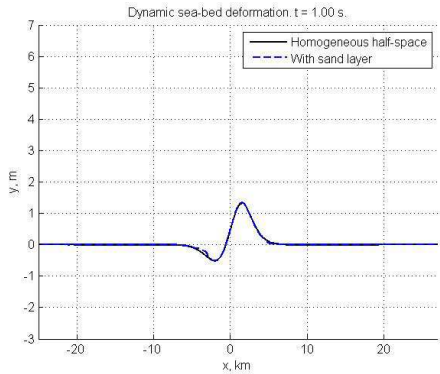}}
  \subfigure[$t = 2.0 s$]%
  {\includegraphics[width=0.49\textwidth]{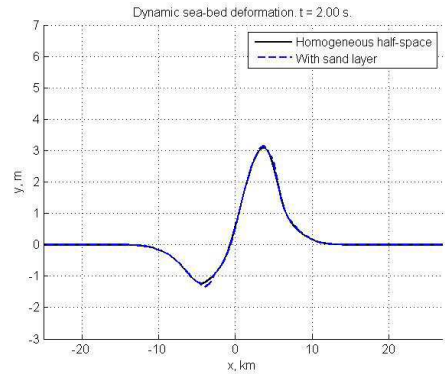}}
  \caption{Dynamic sea-bed displacements at the beginning. The thickness of the sediment layer is $h_s = 150$ m.}
  \label{fig:beginH150}
\end{figure}

\begin{figure}
  \centering
  \subfigure[$t = 3.0 s$]%
  {\includegraphics[width=0.49\textwidth]{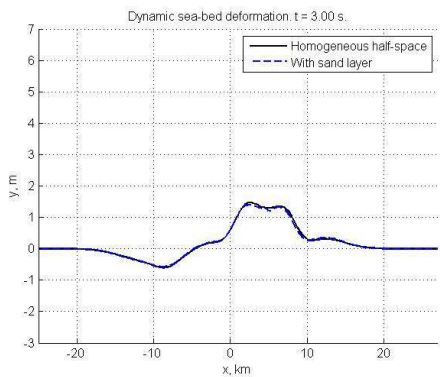}}
  \subfigure[$t = 3.5 s$]%
  {\includegraphics[width=0.49\textwidth]{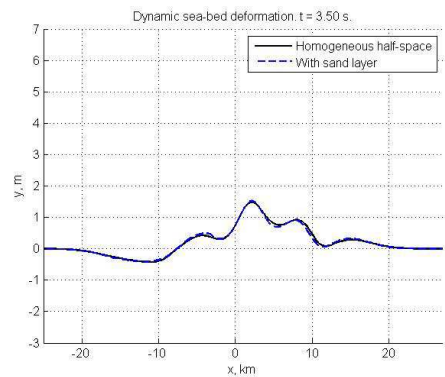}}
  \caption{Dynamic sea-bed displacements. End of the process. The thickness of the sediment layer is $h_s = 150$ m.}
  \label{fig:endH150}
\end{figure}

We performed other computations where the sediment layer thickness was reduced to $150$ m. Results are presented on Figures \ref{fig:beginH150}--\ref{fig:endH150}. In this case, both solutions (homogeneous and inhomogeneous) evolve together and are almost indistinguishable within graphical accuracy. These results suggest to study the dependence of the vertical displacement amplitude on the sediment layer thickness.

\subsection{Sediment amplification factor}

In order to quantify the influence of sediments on the vertical seabed displacements, we introduce a new quantity $\S_a$ that is called the \textit{sediment amplification factor}. We give first the formula for $\S_a$ and then explain our definition.

\begin{definition}
Let us denote by $v_0(x,t)$ and $v_s(x,t)$ the vertical displacements at the free surface in a homogeneous half-space and at the top of the sediment layer respectively\footnote{In the idealized situation of our test cases, it means that we evaluate the vertical displacements at $z = 0$.}. Then, the sediment amplification factor is defined as follows:
\begin{equation*}
  \S_a = \frac{\max\limits_{(x,t)}|v_s(x,t)|}{\max\limits_{(x,t)}|v_0(x,t)|} - 1.
\end{equation*}
\end{definition}

Let us provide some explanations. First of all, it is clear that we compare the values of two extreme amplitudes. The maximum is taken in both space and time, since both processes are not synchronised in time\footnote{We saw on Figures \ref{fig:beginH600}--\ref{fig:endH600} that the homogeneous solution evolves faster since sediments slow down the properties of elastic wave propagation.}. Finally, we substract one because we want the amplification factor to be equal to zero when sediments are absent.

Once this quantity $\S_a$ is defined, a parametric study must be performed. Here we are mainly interested in the dependence on the sediment layer thickness. But it is better to choose a dimensionless quantity. In this problem there are three lengths: the fault length $L$, the fault depth $d$ and the sediment layer thickness $h_s$. It is natural to choose the ratio $h_s/d$ as dimensionless parameter.

\begin{figure}
  \centering
  \includegraphics[width=0.9\textwidth]{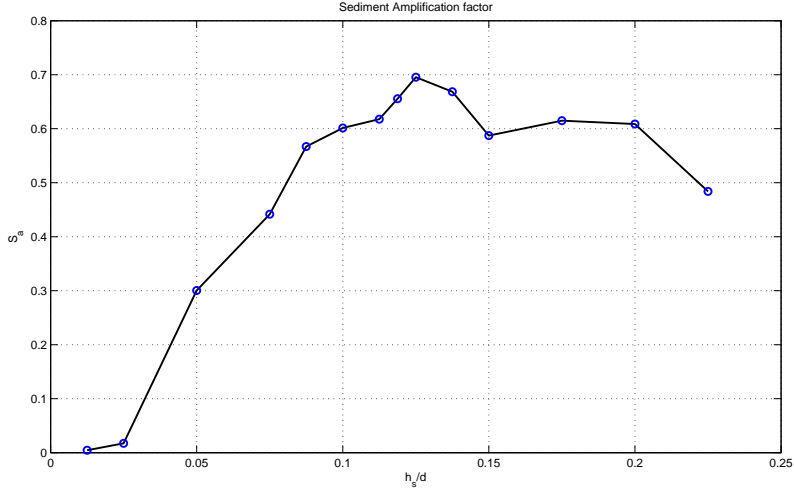}
  \caption{Dependence of the sediment amplification factor $S_a$ on the sand layer relative depth $h_s/d$.}
  \label{fig:sedimpact}
\end{figure}

We performed a lot of computations for different values of $h_s/d$ and obtained the curve shown on \figurename~\ref{fig:sedimpact}. It leads to several comments. On the left, the curve starts from zero and it is expected since the sediment layer disappears at this extremity. Thus, its amplification is equal to zero as well. It is interesting that the amplification factor has a maximum in the vicinity of $h_s/d = 0.12$. It means that there exists an optimal configuration when the sediment layer has its strongest effect. When we gradually increase the dimensionless parameter $h_s/d$ past the maximum, the amplification decreases. In the limit, one has to be careful as $h_s/d \to \infty$. Indeed one approaches an elastic half-space completely filled with sediment and dislocation theory may no longer be appropriate, especially if the material is loose. 

\section{Conclusions and perspectives}\label{sec:conclusions}

In the present paper we investigated the influence of sedimentary layering on displacements due to an earthquake. We showed that there is practically no effect in the case of static deformation. This is to be expected in the framework of our model. Both curves can be superimposed up to graphical accuracy. On the other hand, dynamics makes a big difference. Our computations show that the vertical displacement amplitude can be amplified by a factor up to $1.7$. We point out that there exists some kind of ``optimal'' sediment layer thickness, which provides the biggest amplification factor. Of course, this optimal value depends on various mechanical parameters. It can be estimated in each specific situation by similar numerical techniques.

There is another predictible effect of the sediment layer. It slows down considerably the velocity of propagation of elastic and Rayleigh waves. In our simulations it is reflected by the fact that the maximum amplitude is reached much later than in the homogeneous case.

We introduced a new quantity $\S_a$, that we called \textit{sediment amplification factor}. This dimensionless quantity measures the relative increase of the vertical displacement amplitude with respect to the homogeneous half-space solution. The dependence of this quantity $\S_a$ on the dimensionless thickness of the sediment layer was studied. We showed that there exists an optimal ratio $h_s/d \approx 0.125$ between sediment thickness and depth of the event which provides the biggest amplification factor $\S_a \approx 0.7$.

It is of interest to see how these results apply to the 2004 Indian Ocean Earthquake, which produced a megatsunami
with local runup greater than 30 m at some locations \citep{SynoKong}. According to \figurename~\ref{fig:sedthick9}, we can estimate the sediment thickness $h_s$ in the generation region to be about $3$ km. In \citep{Lay, indiens2}, the centroid depth $d$ was set to $25$ km for all fault subdivisions. If we compute the ratio of these parameters, we obtain $h_s/d = 0.12$. This value approximatively corresponds to the value (see \figurename~\ref{fig:sedimpact}) which provides the maximal sediment amplification factor. The natural question is: is it a coincidence? It is not possible to answer at this stage. Indeed the values we have used for the fault size are not realistic. Most likely there can be an amplification in real situations but not as large.

The overall conclusion of this study is that one may have to revise the initial conditions used in some tsunami simulations. More precisely one has to take care of situations where the generation region contains sediment deposits. Most likely it was the case of the Boxing Day Tsunami of 2004 \citep{Syno2006}. Several researchers had to take unphysically large values of the slip along the fault\footnote{In terms of dislocations, it means the absolute value of the Burgers vector.} in order to generate a significant tsunami wave (see for example \citep{Ioualalen2007}). If one takes sediments into account, this value can be reduced while producing the same wave amplitude.

We finally outline some directions for future research in this field. First of all, the application of these techniques to real world events requires, of course, 3D computations, even if we do not think that it will change qualitatively our results. On the other hand, the fracturing through the sediments should be further investigated since it was conjectured to provide a much bigger amplification factor \citep{Fukao1979, Okal1988}. At the same time, the question of the influence of sediment porosity has not been addressed in the present study and is left for future investigations. We have the feeling that porosity may enhance sea-bed deformations in the near field but it should be checked by thorough computations.

Finally it should be emphasized that the present study has focused on the deformation of the sea-bed. The question of the amplification of tsunami waves is a different one. Even if there is an effect in the near-field, it is not obvious that there is an effect in the far-field. The normal mode theory used by Okal \cite{Okal1988} provides an ideal framework to study both near-field and far-field effects. Moreover it is relatively easy to vary as many parameters as possible (focal depth, rupture velocity, dip angle, slip angle, thickness of sediment layer, fault size, sedimentary structure, etc).



\section*{Acknowledgment}

The authors would like to thank the instigator of this work, Professor Costas Synolakis, for indicating new research directions. Special thanks go to Professor Emile Okal for very helpful discussions and valuable suggestions.

The second author acknowledges the support from the EU project TRANSFER (Tsunami Risk ANd Strategies For the European Region) of the sixth Framework Programme under contract no. 037058 and the support from the 2008 Framework Program for Research,
Technological development and Innovation of the Cyprus Research Promotion Foundation under the Project A$\Sigma$TI/0308(BE)/05. 

\bibliographystyle{alpha}
\bibliography{biblio}

\newcommand{\etalchar}[1]{$^{#1}$}
\begin{thebibliography}{HRR{\etalchar{+}}04}

\bibitem[AR02]{Aki2002}
K.~Aki and P.G. Richards.
\newblock {\em Quantitative Seismology}.
\newblock University Science Books, 2002.

\bibitem[BMR72]{Ben-M}
A.~Ben-Menahem and M.~Rosenman.
\newblock Amplitude patterns of tsunami waves from submarine earthquakes.
\newblock {\em J. Geophys. Res.}, 77:3097--3128, 1972.

\bibitem[DD07]{Dutykh2006}
D.~Dutykh and F.~Dias.
\newblock Water waves generated by a moving bottom.
\newblock In Anjan Kundu, editor, {\em Tsunami and Nonlinear Waves}. Springer
  Verlag (Geo Sc.), 2007.

\bibitem[DD09]{DD09}
D.~Dutykh and F.~Dias.
\newblock Tsunami generation by dynamic displacement of sea bed due to dip-slip
  faulting.
\newblock {\em Mathematics and Computers in Simulation}, In press, 2009.

\bibitem[DDK06]{ddk}
D.~Dutykh, F.~Dias, and Y.~Kervella.
\newblock Linear theory of wave generation by a moving bottom.
\newblock {\em C. R. Acad. Sci. Paris, Ser. I}, 343:499--504, 2006.

\bibitem[Dut07]{Dutykh2007a}
D.~Dutykh.
\newblock {\em Mathematical modelling of tsunami waves}.
\newblock PhD thesis, \'{E}cole {N}ormale {S}up\'{e}rieure de {C}achan, 2007.

\bibitem[EOS{\etalchar{+}}05]{Ergun2005}
M.~Ergun, S.~Okay, C.~Sari, E.Z. Oral, M.~Ash, J.~Hall, and H.~Miller.
\newblock Gravity anomalies of the {C}yprus {A}rc and their tectonic
  implications.
\newblock {\em Marine Geology}, 221:349--358, 2005.

\bibitem[FB76]{Freund1976}
L.~B. Freund and D.~M. Barnett.
\newblock A two-dimensional analysis of surface deformation due to dip-slip
  faulting.
\newblock {\em Bull. Seism. Soc. Am.}, 66:667--675, 1976.

\bibitem[Fuk79]{Fukao1979}
Y.~Fukao.
\newblock Tsunami earthquakes and subduction processes near deep-sea trenches.
\newblock {\em J. Geophys. Res.}, 84:2303--2314, 1979.

\bibitem[FWB06]{Fuller2006}
C.W. Fuller, S.D. Willett, and M.T. Brandon.
\newblock Formation of forearc basins and their influence on subduction zone
  earthquakes.
\newblock {\em Geology}, 34:65--68, 2006.

\bibitem[Gus78]{Gusiakov1978}
V.K. Gusiakov.
\newblock {\em Ill-posed problems of mathematical physics and interpretation of
  geophysical data}, chapter Static displacement on the surface of an elastic
  space (in Russian), pages 23--51.
\newblock VC SOAN SSSR, 1978.

\bibitem[HPHO98]{Hecht1998}
F.~Hecht, O.~Pironneau, A.~Le Hyaric, and K.~Ohtsuka.
\newblock {\em FreeFem++}.
\newblock Laboratoire JL Lions, University of Paris VI, France, 1998.

\bibitem[HRR{\etalchar{+}}04]{Hoogakker2004}
B.A.A. Hoogakker, R.G. Rothwell, E.J. Rohling, M.~Paterne, D.A.V. Stow, J.O.
  Herrle, and T.~Clayton.
\newblock Variations in terrigenous dilution in western {M}editerranean {S}ea
  pelagic sediments in response to climate change during the last glacial
  cycle.
\newblock {\em Marine Geology}, 211:21--43, 2004.

\bibitem[IAK{\etalchar{+}}07]{Ioualalen2007}
M.~Ioualalen, J.~Asavanant, N.~Kaewbanjak, S.T. Grilli, J.T. Kirby, and
  P.~Watts.
\newblock Modeling the 26 {D}ecember 2004 {I}ndian {O}cean tsunami: Case study
  of impact in {T}hailand.
\newblock {\em Journal of Geophysical Research}, 112:C07024, 2007.

\bibitem[KDID07]{KDID}
C.~Kassiotis, F.~Dias, A.~Ibrahimbegovic, and D.~Dutykh.
\newblock A partitioned approach to model tsunami impact on coastal
  protections.
\newblock In A.~Ibrahimbegovic, F.~Dias, H.~Matthies, and P.~Wriggers, editors,
  {\em ECCOMAS Thematic Conference on Multi-scale Computational Methods for
  Solids and Fluids}, pages 134--139, 2007.

\bibitem[LKA{\etalchar{+}}05]{Lay}
T.~Lay, H.~Kanamori, C.~J. Ammon, M.~Nettles, S.~N. Ward, R.~C. Aster, S.~L.
  Beck, S.~L. Bilek, M.~R. Brudzinski, R.~Butler, H.~R. DeShon, G.~Ekstrom,
  K.~Satake, and S.~Sipkin.
\newblock The great {S}umatra-{A}ndaman earthquake of 26 {D}ecember 2004.
\newblock {\em Science}, 308:1127--1133, 2005.

\bibitem[Mei94]{Mei1994}
C.C. Mei.
\newblock {\em The applied dynamics of ocean surface waves}.
\newblock World Scientific, 1994.

\bibitem[MS71]{Mansinha1971}
L.~Mansinha and D.~E. Smylie.
\newblock The displacement fields of inclined faults.
\newblock {\em Bull. Seism. Soc. Am.}, 61:1433--1440, 1971.

\bibitem[NSS{\etalchar{+}}05]{indiens2}
S.~Neetu, I.~Suresh, R.~Shankar, D.~Shankar, S.S.C. Shenoi, S.R. Shetye,
  D.~Sundar, and B.~Nagarajan.
\newblock Comment on ``{T}he {G}reat {S}umatra-{A}ndaman {E}arthquake of 26
  {D}ecember 2004''.
\newblock {\em Science}, 310:1431a--1431b, 2005.

\bibitem[Oka85]{Okada85}
Y.~Okada.
\newblock Surface deformation due to shear and tensile faults in a half-space.
\newblock {\em Bull. Seism. Soc. Am.}, 75:1135--1154, 1985.

\bibitem[Oka88]{Okal1988}
E.~Okal.
\newblock Seismic parameters controlling far-field tsunami amplitudes: A
  review.
\newblock {\em Natural Hazards}, 1:67--96, 1988.

\bibitem[OS03]{OkalSyno2003}
E.A. Okal and C.E. Synolakis.
\newblock A theoretical comparison of tsunamis from dislocations and
  landslides.
\newblock {\em Pure and Applied Geophysics}, 160:2177--2188, 2003.

\bibitem[SB06]{Syno2006}
C.E. Synolakis and E.N. Bernard.
\newblock Tsunami science before and beyond {B}oxing {D}ay 2004.
\newblock {\em Phil. Trans. R. Soc. A}, 364:2231--2265, 2006.

\bibitem[SK06]{SynoKong}
C.E. Synolakis and L.~Kong.
\newblock Runup measurements of the {D}ecember 2004 {I}ndian ocean tsunami.
\newblock {\em Earthquake Spectra}, 22:S67--S91, 2006.

\bibitem[SLCY97]{SLCY1997}
C.~Synolakis, P.~Liu, G.~Carrier, and H.~Yeh.
\newblock Tsunamigenic sea-floor deformations.
\newblock {\em Science}, 278:598--600, 1997.

\end{thebibliography}
\end{document}